\documentclass[12 pt]{article}
\usepackage[utf8]{inputenc}
\usepackage{authblk}
\usepackage[sectionbib]{natbib}
\usepackage{chapterbib}
\usepackage{epigraph}
\usepackage{hyperref}
\usepackage{graphicx}
\usepackage{mathtools}   % loads »amsmath«
\usepackage{amsfonts}
\usepackage{appendix}
\usepackage{caption}
\usepackage{orcidlink}
\graphicspath{ {./figures/} }
\hypersetup{
    colorlinks=true,
    linkcolor=blue,
}
\providecommand{\keywords}[1]
{
  \small	
  \textbf{\textit{Keywords---}} #1
}

\begin{document}

\title{Geometrogenesis in GFT: an analysis}

\author[1,2]{Álvaro Mozota Frauca}
\affil[ ]{alvaro.mozota@uab.cat, \orcidlink{0000-0002-7715-0563} \href{https://orcid.org/0000-0002-7715-0563}{https://orcid.org/
0000-0002-7715-0563}}
\affil[1]{Department of Philosophy. Universitat Aut\`onoma de
Barcelona, Building B Campus UAB 08193 Bellaterra (Barcelona), Spain}
\affil[2]{LOGOS, University of Barcelona, Department of Logic, History and
Philosophy of Science, Carrer de Montalegre 6 08001 Barcelona, Spain}

\maketitle

\begin{abstract}
In this paper I introduce the idea of geometrogenesis as suggested in the group field theory literature and I offer a criticism of it. Geometrogenesis in the context of GFT is the idea that what we observe as the big bang is nothing else but a phase transition from a non-geometric phase of the universe to a geometric one which is the one we live in and the one to which the spacetime concepts apply. GFT offers the machinery to speak about geometric and non-geometric phases, but I argue that there are serious conceptual issues that threaten the viability of the idea. Some of these issues are directly related to the foundations of GFT and are concerned with the fact that it isn't clear what GFT amounts to and how to understand it. The other main source of trouble has to do with geometrogenesis itself and its conceptual underpinnings as it is unclear whether it requires the addition of an extra temporal or quasitemporal dimension which is unwanted and problematic.
\end{abstract}


Group field theory (GFT) is an approach to quantum gravity that offers the resources to speak about different phases our universe could be at: geometrical ones in which the standard spatiotemporal description applies and non-geometrical ones which would be radically different. This allows us to speculate about whether we should replace the big bang picture with a picture in which the universe underwent a phase transition from a non-spatiotemporal phase to our spatiotemporal one. In the context of GFT there is a proposal for such a phase transition, called geometrogenesis, which I analyze in this paper. In particular, I will find that there are serious issues that make the proposal problematic.  

I will start in section \ref{GFT_101} by giving a basic overview of the GFT approach and by highlighting that it lacks a clear formulation and that there are conceptual issues that need to be addressed. These issues will have an impact on the formulation of the geometrogenesis idea, as they will make it difficult to form a clear picture of how this proposal is to be implemented in the GFT formalism.

In section \ref{GFT_condensates_cosmo} I briefly review the most extended family of GFT cosmological models, which are based on the idea that the quantum state of the universe in its geometric phase can be described by a condensate state. This picture is interesting on its own, and can be considered independently of geometrogenesis, i.e., there are cosmological models based on condensate states that do not need to appeal to geometrogenesis. I explain how dynamics is defined in this class of models and I anticipate that this definition of the dynamics is different from the one that is postulated to drive geometrogenesis, and hence, that there will be some tension between the two.

Then, in section \ref{Phase_transition_general} I introduce the notion of phase transition in general and its relation with the renormalization group flow before getting into the concrete version of geometrogenesis proposed in the context of GFT in section \ref{Geometrogenesis_GFT}. I will argue that the proposal isn't very clear and that at some moments it seems to need the addition of an unwanted extra temporal dimension. In this sense, I conclude in section \ref{conclusions} that the idea of geometrodynamics, in the way it is formulated and in the way it relies on GFT, is not compelling.

\section{GFT in a nutshell}\label{GFT_101} 

GFT is an approach to quantum gravity that can be motivated by some formal features of a few different approaches to quantum gravity\footnote{See for instance \cite{Oriti2007} for an explanation of these connections between approaches and with GFT.}. For instance, the form of some spin foam models suggests taking them as terms on a Feynman expansion, and one can see that a way of generating such an expansion is by formulating a field theory that is defined not on a spacetime manifold but on a group manifold. In this way one arrives at the basic idea of GFT: that the fundamental theory underlying several approaches to quantum gravity, and underlying general relativity too has to be a quantum field theory defined on a group manifold.

From a formal point of view a GFT is defined\footnote{See for instance the definitions in \cite{Freidel2005,Oriti2012}.} by specifying a group manifold $G$, a field $\varphi$ defined on it, and an action for this field $S[\varphi]$. This allows defining quantities like partition functions and expectation values by means of path integrals:
\begin{equation} \label{GFT_partition_function}
Z=\int \mathcal{D}\varphi  e^{-S[\varphi]} 
\end{equation}
\begin{equation} \label{GFT_expectation_value}
\langle f\rangle=\int \mathcal{D}\varphi f(\varphi) e^{-S[\varphi]} \, ,
\end{equation}
where $\mathcal{D}\varphi$ represents some appropriate measure on the space of field configurations on $G$ and $f$ is some functional of the field. By conveniently expanding these expressions in power series of the relevant parameters in the action one recovers expressions from spin foam models and other discrete models\footnote{This connection between Feynman diagrams and spin foam amplitudes was first discussed in \cite{Reisenberger2001}.}.

The formal definition above is a little bit wanting from the conceptual point of view, in the sense that it seems to amount just to the definition of some computational rules useful for computing quantities that can be interesting from the point of view of other approaches. Moreover, it doesn't provide a quantum theory in the standard sense, i.e., it doesn't define the usual structures like Hilbert spaces, operators, and so on. 

Despite this, in the GFT literature one also finds a `canonical' version of GFT, i.e., one defined using states in a Hilbert space. As far as I know, and contrary to the way it is presented in the literature\footnote{Consider for instance the discussion of the canonical version of GFT in \cite{Gielen2013a}.}, there is no proof in the literature showing the equivalence of both formulations of GFT\footnote{One can at most do some heuristic comparison with the case of standard QFT, but while in the case of QFT one can show the equivalence between covariant and canonical formalisms, the ways of proving this rely on the spacetime structures and Hamiltonian dynamics of standard field theories, which are not available for the case of GFT.}, and hence I will take the canonical version to be just postulated. This canonical version provides the basis on which the talk of condensate states for describing the geometric phase of the universe is based.

%However, in the GFT literature it is claimed that this definition does define a quantum theory in the standard sense. Here I won't argue against this claim and I will just introduce the `canonical' version of GFT as the version defined using states in a Hilbert space. In this sense, I take that the canonical version of GFT is just postulated at least until a proof of the equivalence is provided. 

In this version of GFT, the GFT Hilbert space is taken to be the Fock space built for the group field theory\footnote{Take \cite{Gielen2013a} as a representative of this way of defining canonical structures for GFT.}. For GFT's aiming to recover 4-dimensional gravity the group manifold usually consists of four copies of SU(2) and hence a state in this Fock space can be characterized as a superposition of different numbers of excitations or `particles', each of them described by four group elements. This is just as in the Fock spaces used in standard quantum field theory states describe a number of particles, each of them characterized by its position, or, equivalently, its momentum.

The GFT Fock space is similar in some aspects to the Hilbert spaces of other approaches to quantum gravity, especially loop quantum gravity (LQG) and models based on simplicial decompositions of space\footnote{I refer the reader to \cite{Rovelli2004} for an introduction of LQG and to \cite{Rovelli2015} for a presentation of LQG that discusses its relationship with triangulations and other decompositions of spacetime. The relationship between GFT and LQG is discussed in detail in \cite{Oriti2016b}.}. This suggests interpreting each GFT quantum as representing an `atom' of space. For the 4-dimensional theory, these atoms have been suggested\footnote{See \cite{Oriti2017} for instance.} to be tetrahedra, and the four group elements describing each atom contain geometric information associated with the four faces of the tetrahedron. In this sense, one can see GFT states as containing a bunch of tetrahedra, and for some particular states for which some further conditions apply one can even argue that these states represent a 3-dimensional space composed of tetrahedra glued together\footnote{For the details about which states correspond to `glued' tetrahedra see the discussion in \cite{Chirco2019}.}.

In this canonical version of GFT it is postulated that the dynamically relevant states, i.e., the states which contain information about evolution, are the ones that satisfy a constraint equation like the following:
\begin{equation}\label{constraint_GFT}
\frac{\delta S[\hat{\varphi}]}{\delta \hat{\varphi}(g)}\vert \psi \rangle=0 \, .
\end{equation}
This equation makes a connection with the covariant version, as it explicitly depends on the group field theory action and it can be heuristically motivated\footnote{See for instance \cite{Gielen2016}.} as being the quantum version of the Euler-Lagrange equations for the field theory. Alternatively, it is common to see in the literature the claim that the dynamically relevant states are the ones that satisfy a series of operator equations similar to the Schwinger-Dyson equations from field theory. These two views are presented as equivalent, although, to the best of my knowledge, no proof of equivalence has been given in the literature\footnote{For discussions of these equations and their relation with the constraint equation see \cite{Gielen2013a, Oriti2017}.}. In any case, these definitions of the dynamics seem to be built just on some heuristics and one lacks a rigorous connection with GFT as built in terms of path integrals.

At the time of interpreting these supposedly dynamical states we find a problem analogous to a well-known problem that affects canonical approaches to quantum gravity: the problem of time. In canonical approaches to quantum gravity, and in this canonical version of GFT, one doesn't have a standard dynamical equation describing how quantum states evolve with respect to a time parameter but just `timeless' equations\footnote{The most famous of these timeless equations is the Wheeler-deWitt equation of quantum geometrodynamics, although any approach to quantum gravity built by applying canonical quantization methods to general relativity will suffer from the problem of time and have a timeless equation, which corresponds to the quantum version of the Hamiltonian constraint. This includes LQG.} that seem to have as solutions something like superpositions of (possibly with some caveats) 3-geometries. The interpretation of these seemingly timeless equations and states has been widely debated in the quantum gravity and philosophy of physics communities. I refer the reader to the literature for more detailed discussions of the problem and the way it affects canonical approaches to quantum gravity\footnote{The classical reviews of the problem of time are \cite{Kuchar1992,Isham1993} and the recent book \cite{Anderson2017} offers a complete survey of it. The quantum gravity community generally considers that the approaches affected by the problem of time nevertheless define meaningful theories, as defended for instance in \cite{Rovelli2004,Kiefer2012,Rovelli2015}. See also \cite{Gryb2016,Chua2021,MozotaFrauca2023} for some critical views.}.

In the context of GFT it has been proposed\footnote{This is explicit in \cite{Oriti2017, Marchetti2021}.} to deal with this problem by introducing additional degrees of freedom to the theory and by implementing a relational definition of the dynamics. That is, to introduce at least a variable $\phi$ which would play the role of a clock or time variable\footnote{In relational approaches there is a problematic conflation of both concepts.}. In particular one expands the base manifold by adding one real variable $\phi$ and applies the same procedure\footnote{Some alternative ways of introducing $\phi$ are available, but here I am just following the GFT literature.} above but for the expanded manifold. This means that GFT particles are now tetrahedra, which on top of their geometrical information also carry a value of $\phi$. This is seen in the GFT literature as each tetrahedron carrying a clock. For the dynamics, one appropriately modifies the GFT action and the constraint equation \ref{constraint_GFT}.

For a one-particle state one can easily see how this proposal would in principle work. Suppose that there is a one-particle state that is a solution to the modified version of equation \ref{constraint_GFT}. One could read the wavefunction $\psi(g_1,g_2,g_3,g_4,\phi)$ as describing a universe formed by just one tetrahedron with a shape and size which is evolving in time $\phi$\footnote{Even in this case one could raise worries such as that prima facie nothing assures you that the evolution in $\phi$ is unitary.}. However, when one has several particles, each carries its own clock, and postulating that one is the preferred time variable seems ad hoc. Moreover, one can have superpositions with terms with different numbers of particles, for which $\phi$ cannot serve as a time variable for the state. In the GFT literature this kind of state is known to be problematic, and sometimes they are included in the non-geometric category. In my opinion, what these states show is that the relational resolution proposed is deeply flawed\footnote{For completeness, let me mention that there are several ways of implementing a relational resolution to the problem of time of reparametrization invariant models. Some authors (see \cite{Rovelli1991a}) prefer a Heisenberg-like picture in which evolution is to be found in the operators that can be defined on the space of states satisfying the dynamical conditions. Similarly to what discussed in the Schr\"odinger picture, the relational strategy could be argued to work for the one-particle subspace, while it fails when one considers the full Hilbert space, as can be seen by rehearsing a similar argument to the one just given. For more recent discussions of relational resolutions of the problem of time I refer the reader to \cite{Hohn2021} and references therein.}.

%. %In any case, my main objection in this paper to the picture of geometrodynamics in GFT is mostly independent of this point as will be clear later on.

For the discussion in this paper, it is important to emphasize that GFT allows for the existence of non-geometric states which are indispensable for the claim that what we know as the big bang singularity was indeed a phase transition from a non-geometric phase to a geometric one. These non-geometric states would correspond to states in which the atoms of space aren't nicely glued to each other in a way that approximates a smooth manifold or states in which the `times' of the different atoms are widely different, in a way that cannot be associated with a picture in which they march in time. These non-geometric states would be the starting point for geometrogenesis. The ending point would be some condensate state, which I introduce in the next section.

\section{GFT condensates and cosmology}\label{GFT_condensates_cosmo}

For cosmological models inspired by GFT, condensate states are extensively used, and it is also the case that geometrogenesis is conceived as a phase transition very similar to the phase transition that leads to the formation of a condensate in a condensed matter system. The use of condensates for describing cosmological settings is motivated by the observation that we are interested in describing large expansions of space that are highly homogeneous. This leads to the idea that we need states formed by a huge number of very symmetric tetrahedra. Condensate states offer the machinery needed for describing in a simple and powerful way states with a great number of particles in very similar states, and for this reason they were introduced in GFT cosmology\footnote{See \cite{Pithis2019,Gabbanelli2020} for recent reviews.}.

A condensate state\footnote{For a general introduction to Bose-Einstein condensates I refer the reader to \cite{Rogel-Salazar2013}.} is a state for a bosonic system in which every or nearly every particle is on the ground state. In particular, we can approximate the $N$-body wavefunction as a product state of $N$ identical states:
\begin{equation} \label{condensate_state_N_part}
\vert\Psi \rangle= \vert \psi \rangle\otimes \vert \psi \rangle\otimes ....\otimes  \vert \psi \rangle \, .
\end{equation} 
This approximation is a great simplification that allows describing the quantum state of a system of many particles in terms of just one one-particle wavefunction $\psi(x)$. That is, instead of having to deal with a complicated many-particle state living in a high-dimensional configuration space one just treats the state as being very symmetric and describable in terms of the wavefunction of just one particle, as it represents the quantum state of any of the particles in the condensate. For the case of GFT, taking a condensate implies assuming that all the tetrahedra are nearly in the same state and that by knowing the geometry of one of them one can know the geometry of all of them to a good degree of approximation.

There is a further step in the approximation which is to describe condensates using coherent states. These states aren't eigenstates of the number operator, i.e., they don't contain a fixed number of particles, but they can be shown to be a good approximation for states with a large number of particles. Coherent states on the other hand are eigenstates of the field operator, which facilitates their formal manipulation. In this sense, coherent states are the tools used in GFT cosmology for describing states formed by a great number of very similar tetrahedra in terms of a simple effective wavefunction $\psi(g,\phi)$, which would be representative of the wavefunction of any of the tetrahedra.

When imposing the dynamics defined by the constraint equation \ref{constraint_GFT} or by the Schwinger-Dyson equations to coherent states, and by applying a number of approximations, one obtains a dynamics for the effective wavefunction $\psi(g,\phi)$. In this way, the dynamics of a system of many particles has been reduced from a complicated dynamics with many degrees of freedom to a dynamics that is in effect like the one of one particle. Notice however that the effective dynamics is not the dynamics of a single particle, as it generally includes collective effects from the fact that we are dealing with a great number of particles\footnote{In the condensed matter case, this effective equation is known as the Gross-Pitaevskii equation. I refer the interested reader again to \cite{Rogel-Salazar2013} for an introduction to condensates in condensed matter physics.}. 

At the time of interpreting solutions to this effective equation, the same relational attitude towards the problem of time I mentioned in the previous section is held\footnote{See again \cite{Marchetti2021}.}, that is, the parameter $\phi$ is effectively considered as a time parameter in which the effective wavefunction $\psi(g,\phi)$ evolves. I won't repeat my reservations towards this interpretation here\footnote{Let me also insist that these reservations also apply if one adopts a Heisenberg picture standpoint and seeks evolution in operators and not in states.}, but I will note that it is just because of the extremely symmetric properties of the coherent quantum state that this is possible. For our discussion of geometrogenesis it is important to remark that dynamics is also defined by means of equation \ref{constraint_GFT} and is taken to be evolution with respect to the $\phi$ parameter.

Finally, let me mention that in condensed matter systems the effective wavefunction is sometimes interpreted in an interesting way that seems to have been exported to GFT. By applying the Born rule one could argue that the modulus of the effective wavefunction, $\vert \psi (x) \vert^2$, represents something like the probability of finding a particle in position $x$. However, in condensed matter systems $\vert \psi (x) \vert^2$ is taken to be an actual density of particles at position $x$. In this sense, $\vert \psi (x)\vert ^2$ would describe the macroscopic properties of the condensate. A heuristic justification for this is that for thermodynamic systems like the condensates we are in the limit of large $N$, and here we expect frequencies to match probabilities. In other words, we take $\psi$ to describe the overall distribution of particles in the condensate instead of being a property of each particle. In this sense, one could say that $\psi$ has become a classical field\footnote{The field $\vert \psi (x) \vert^2$ is now interpreted as a density field and the phase of $\psi$ is a field that will describe the velocity of the condensate.}, and the predictions made using this classical field interpretation agree with the experimental results\footnote{These results include phenomena like superfluidity and superconductivity.}. From a rigorous point of view, the identification of $\vert \psi (x) \vert^2$ as a macroscopic classical variable is not fully justified unless one deals with the interpretative issues of quantum mechanics and the measurement problem. But for practical applications, this identification is a useful one.

The same kind of identification seems to be in play in the GFT cosmology literature where it seems that there is one interpretative jump from the `microscopic', quantum description of quanta of space to the effective description given in terms of a classical field $\psi(g,\phi)$. This field can be interpreted as giving the number or the density of tetrahedra with geometry given by some group variable $g$ at time $\phi$. By summing up all the tetrahedra and geometries one is able to get an effective scale factor for the universe $a(\phi)$. In this way GFT cosmology aims to connect with observational cosmology, where objects like $a(\phi)$ can in principle be observed. In particular, GFT cosmology generically predicts bounce scenarios, i.e., scenarios in which the universe underwent a contracting epoch before the expanding one in which we currently live. For our discussion of geometrogenesis we won't need any more detail about the GFT cosmological model and I refer the interested reader to the literature\footnote{See for instance \cite{Oriti2016a,Gielen2018,Gielen2019,Pithis2019,Gabbanelli2020}.}.

For the discussion of geometrogenesis it is interesting to note that the condensate picture is taken to be an approximation to the real state of the universe in the geometric phase. There are some important features of a geometric state that are missing from the condensate state, such as the information about the way the different tetrahedra are glued to each other to form a space composed of pieces. However, one can hope that condensate states capture well the cosmologically relevant geometric information, namely the scale factor of the universe. 

The idea of the condensate being an approximation is also interesting when one wants to speculate about geometrogenesis. According to the proponents of geometrogenesis, as we go back in time the condensate approximation should become worse and worse, until we reach the moment of geometrogenesis, which would delimit the instant at which the geometric picture and approximation stops making sense. Let me mention that there are some arguments for supporting this claim, as some of the approximations in play at the time of deriving the effective equations seem to break down for small-volume regimes\footnote{In particular, these approximations cease to be valid when there are big quantum fluctuations. I refer the reader to \cite{Marchetti2021a} for a discussion of quantum fluctuations and the way they jeopardize the bouncing picture. I am thankful to an anonymous reviewer for bringing up this point to me.}.

Note here an important difference with the way approximations are believed to be broken in related cases. While in classical cosmology one can rehearse an argument that something peculiar is happening at the big bang based on the fact that there is a singularity, from the effective Friedman dynamics in the GFT condensate model the dynamics is perfectly smooth and sensible: instead of reaching a point of infinite curvature and no size one generically finds a bounce, i.e., that before the current expanding phase of the universe there was a contracting one until a minimal scale factor was reached. From the perspective of GFT, in order to look for signals of something unusual happening, one has to look at a deeper level and analyze the approximations that are in play. That is, by looking just at the effective equations one cannot tell whether the approximation breaks down or not, contrary to the alleged case of other effective theories. 

In this sense, to decide whether GFT predicts a bounce scenario or something like a geometrogenesis one needs to be careful and study the approximations in play. In this paper I will be taking seriously the possibility that the geometric approximation breaks down and that one needs to consider the option of a geometrogenesis. However, for doing so, one needs to face deep conceptual issues at the time of understanding the way GFT aims to be a meaningful physical theory, as I highlighted in the previous section. That is, for generic states it is not clear to what extent the equation \ref{constraint_GFT} (or the Schwinger-Dyson equations) represents a dynamics or whether it makes sense to consider the clock variable $\phi_i$ associated with some of the tetrahedra to play the role of a time variable. In this sense, we find that the condensate idea allows for tractable models, but it is based on a theory that is not well understood and for which one can have reasonable reservations. Therefore, it will be problematic to claim that there is a phase transition if there is no control over the putative base theory.

\section{Phase transitions} \label{Phase_transition_general}

In the previous two sections I have introduced GFT and the condensate states and, despite the serious conceptual troubles that there are within this approach, I hope to have made clear why in this approach one may want to speak about a phase transition between a non-geometric and a geometric phase, i.e., a geometrogenesis. For conceptual clarity, in this section I make a small digression from GFT to discuss phase transitions in general. In particular, I aim to clarify the role of the renormalization group flow in the analysis of phase transitions and how the parameters in a phase diagram represent something external to the system under study. For a more in-depth introduction to phase transitions and the philosophical issues related to them I refer the reader to \cite{Menon2013}.

\begin{figure}[h]%
\centering
\includegraphics[width=0.3\textwidth]{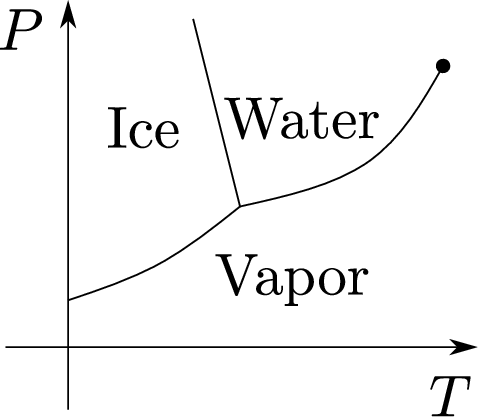}
\caption{\label{phasediagram} Schematic representation of the phase diagram of water. The parameters $T$ and $P$ (temperature and pressure) are parameters that can be changed externally to the system and which fix which phase obtains. The dot represents the critical point. A phase transition would be represented by a line in this diagram joining an initial and a final state in different phases.}
\end{figure}

In thermodynamics, a system is said to have different phases if for different values of its thermodynamic properties, e.g. for different temperatures, it behaves qualitatively differently. The most common example is water: depending on its temperature it can behave as a solid, as a liquid, or as a gas. This can be represented in a phase diagram: this is a diagram that represents the different values the thermodynamic properties of a system can have and it is divided into different regions, which correspond to the different phases. In figure \ref{phasediagram} I schematically represent the phase diagram of water as an example of a phase diagram. 

Notice that in thermodynamics we deal with equilibrium states, so there is no temporal evolution associated with the points of the phase diagram. Now, the thermodynamic properties of a system can be modified\footnote{This can be seen as a process produced by us adjusting some knobs in a lab, or as natural processes happening when systems are not isolated and interact with an environment that changes.} and it can be the case that the system that started in one given phase ends up in a different one. An example of this process is what happens when we heat up ice until it melts. This process is called phase transition and is represented in the phase diagram by a flow connecting two points in different regions, that is, in different phases. As it is standard in thermodynamics, a trajectory in the phase diagram does not carry information about the duration of the process, as we many times are interested in questions like what the properties of the system would be if cooled down to a certain temperature and not that much in the time it takes for this cooling to happen\footnote{Of course, in certain contexts one is interested in making predictions about the duration of a process, but when studying phase transitions one generally does not worry about this context-sensitive predictions.}. In this sense, the power of thermodynamics is that it predicts that water will freeze if cooled down below $0 ^\circ C$, independently of whether this cooling down process is quick or slow or the way this cooling down is achieved.

It should be clear that phase transitions as they are usually understood in condensed matter physics are temporal processes, that is, processes happening in space and time. This will be in contrast with some other `processes' that are not spatiotemporal nor physical, as they will be just computational processes. Take for instance the example of an order-by-order perturbative approximation for a given physical problem, say, the estimation of the ground state energy $E$ of an atom. Perturbative calculations will give us a sequence of estimations $E^{(0)},E^{(1)},E^{(2)}...$ that will, if we are employing the right techniques, be approaching the physical value of $E$. This computational process allows us to make physical predictions and it captures some physical phenomena, but the sequence $E^{(0)},E^{(1)},E^{(2)}...$  clearly does not represent any physical process. That is, the energy of the atom in the world is not undergoing such changes as it is just $E$. The distinction between physical processes and abstract or computationally useful `processes' will be important for the discussion in this article. The latter can have certain physical significance in the sense that they can be used for making certain predictions like the value of $E$ in the example, but I want to emphasize that they are not physical in the strong sense that a piece of ice melting is a physical process. In a different context, one would define a physical process as a process happening in space and time, but this definition is not satisfactory if we want to consider the possibility of a spacetimeless theory or regime. But even in this case, one will also want to distinguish between processes that are physical and just computationally useful `processes'. I leave the discussion of non-spatiotemporal but physical processes for the next section, but before that I will introduce the renormalization group flow as it is commonly used, and I will argue that it is not a physical process, that is, it generally does not represent a process in space and time. Even if in the quantum gravity context the definition of physical process is harder to pin down, if the renormalization group flow is considered not to be physical in the condensed matter or high-energy physics contexts, if in a quantum gravity situation one wants to claim it is physical, then some further argument seems to be needed.

Coming back to phase transitions, there are two types of them: first-order and continuous\footnote{This classification of phase transitions is standard in the condensed matter literture \citep{Binney1992,Goldenfeld1992,Cardy2015}.}. First-order phase transitions are transitions like the ice-water or water-vapor phase transitions we are used to, while continuous phase transitions are more exotic, as they happen just at critical points. If we take a look at the phase diagram of water in figure \ref{phasediagram} we see how there is a boundary between the liquid and gaseous phases that just ends at a certain point. This point is the critical point of water and it is the point at which the two phases of water become the same, i.e., from that point on one can speak about a supercritical fluid and not about liquid and gas. Critical points and continuous phase transitions occur for a wide variety of systems, and have interesting properties that have been extensively studied by the condensed matter physics community.

There are two properties of continuous phase transitions and critical points that are relevant to our discussion. First, critical phenomena, i.e., interesting physical phenomena happening at critical points are many times signaled by singular behaviors of thermodynamic properties. The fact that at the big bang we have singularities at certain geometric properties could be taken to support the view that the big bang can be identified with a phase transition and that it is a continuous one\footnote{I am thankful to an anonymous reviewer for bringing this point up.}. Indeed, \cite{Oriti2014} mentions singularities as `hints' that would be supporting the ideas that new, non-spatiotemporal physics is needed and that phase transition may be associated with spacetime singularities. \cite{Oriti2021} mentions critical values when discussing geometrogenesis, so I believe that it is reasonable to assume that he has in mind a continuous phase transition when discussing geometrogenesis\footnote{A completely explicit argument for the claim that the big bang singularity is a critical point can be found in \cite{Mielczarek2017}, which Oriti cites in \cite{Oriti2021}.}. Despite this, I believe that the idea of geometrogenesis does not commit one to take the phase transition to be continuous, and my arguments in this article won't depend on the type of phase transition one takes geometrogenesis to be. The second property of continuous phase transitions that is relevant to our discussion is that at critical points systems become scale-invariant, which roughly speaking means that they look the same no matter at which scale you look at them.

To study the scale dependence of the properties of a system more formally, one uses a family of methods known as the renormalization group\footnote{For technical discussions of this see \cite{Binney1992,Goldenfeld1992,Cardy2015}.}. These methods allow identifying critical points and studying the phase diagram of condensed matter system. Roughly speaking, when we apply the renormalization group to a condensed matter system, we zoom out to get a coarser grained picture of the system, that is, we blur the short scale details to focus on the large scale. The fact that critical points are scale invariant implies that when we apply the renormalization group to the state represented by such a point, we obtain just the same point, as the system just looks the same from every scale, and zooming out does not affect it. When a system is at some different point on the phase diagram and we zoom out, the properties of the zoomed out version will look different. For example, for a gas we would say that the zoomed out version has a different temperature and pressure. The renormalization group then defines a flow on phase space: it connects each point with others that can be seen as representing zoomed out versions of the system represented by the original point. Now, all the points in the same phase show the same behavior under this flow, that is, they all flow toward/away from the same fixed points\footnote{This analysis of the phases of a system in terms of fixed points is of extended use in the condensed matter literature \citep{Binney1992,Goldenfeld1992,Cardy2015}.}. Therefore, studying the renormalization group flow allows one to identify different phases and the boundaries in a phase diagram, that is, the points at which phase transitions (first-order or continuous)\footnote{Notice that the renormalization group can be used for inferring properties also of first-order phase transitions, and therefore I won't take that discussing the renormalization group flow of a system implies that one is interested just in the case of continuous phase transitions. For the case of geometrogenesis this means that using renormalization group methods for understanding it does not commit one to believing that it is a critical phase transition.} happen.

Let me illustrate this with a paradigmatic example, the case of a magnetic material with no external field\footnote{This example is discussed in practically every introduction to phase transitions and the renormalization group. See for instance the discussions in \cite{Binney1992,Goldenfeld1992,Cardy2015}.}. One can think about this material as a collection of atoms that carry small magnets, and the macroscopic magnetization of the material is obtained just by taking into account the contribution of every small magnet. This system can have two different phases: a ferromagentic phase in which even in the absence of an external field there is some ordering of the magnets that gives rise to a non-zero magnetization, and a paramagnetic phase, in which there is no such ordering and magnetization. If we know the state of every magnet of such material we can coarse-grain to infer in which phase the material is. If the system is in a ferromagnetic state it will have a majority of atoms with their magnets aligned, and a minority not aligned. When we coarse-grain, this minority gets diluted, and the system looks more ordered and the magnets more aligned. For this reason, the flow of the renormalization group flow takes us away from the disordered phase and from the phase transition and leads toward total order (zero temperature). Conversely, in the paramagnetic case, disorder is dominant, and when one zooms out the contributions of the few aligned magnets get diluted and the zoomed out version looks even more disordered (warmer). In between these two cases there is the critical point, that strikes the balance between aligned and not aligned magnets and remains invariant (and at the critical temperature) under this zooming out process. 

Importantly for our discussion, notice that this zooming out process is not a physical process but just a computational technique. Let me insist that nothing is changing in the state of our magnet. We were given information about the state of the magnet as it is, we took pen and paper (or more likely a computer), computed how the coarse-grained version would look like, and determined whether the system was in its paramagnetic phase, in its ferromagnetic phase or at its critical point. But at no moment we have touched the physical magnet and it has remained just the same. One can use renormalization group techniques for computing more interesting and complex properties of condensed matter systems near critical points, but the analysis and interpretation of the renormalization group flow is that is an abstract flow and not a process that is actually happening to a system. Of course, one can (for certain systems) have situations in which the evolution of a system is such that it matches with that of the renormalization group flow, but this is not necessarily the case. For instance, in the magnetic system example, we can have situations in which we cool down a ferromagnet or heat up a paramagnet and this would be represented by trajectories that coincide with the renormalization group trajectories. But we could also heat up a ferromagnet, cool down a paramagnet, first heat up and then cool down a paramagnet, and, of course, heat up a ferromagnet to a temperature above its critical temperature so that it becomes a paramagnet. To insist, condensed matter phase transitions, and changes of state in general are physical processes happening to physical systems, while the renormalization group flow is a useful tool for making predictions for such systems, even if the systems never evolved following the flow of the renormalization group.

There is a different version of the renormalization group flow which is the one that is used in high-energy physics\footnote{For introductions to renormalization in high-energy physics I refer the reader to \cite[Chap. III.1]{Zee2010}, \cite{Butterfield2014a}, and \cite{Williams2018}.}. In this context, the flow is a flow in a space of theories rather than a flow in a phase space which describes systems with different thermodynamic properties. The flow in this case connects theories that have different cut-offs, i.e., different regimes of validity, but which agree in their predictions for the phenomena in the regions of validity they share. For instance, in quantum electrodynamics the renormalization group flow relates two different theories, one with a high-energy cut-off and one with a lower one, and both theories can be used for describing the same phenomena, e.g. the interactions of photons and electrons, at low energies. In this sense, the renormalization group flow does not represent any physical process but it is just a flow in an abstract space of theories that does not affect the physical content of these theories. That we can use a theory $T_1$ or a theory $T_2$ for describing the same phenomenon $P$ at certain conditions doesn't a priori tell us anything about whether the system described by such theories would experience a phase transition if the conditions at which the system are changed. For this reason, we can say that this kind of renormalization group flow isn't directly related to the thermodynamic phases of a system. The historical motivation for introducing the flow in high-energy physics was more technical than physical, as cut-offs were introduced in order to define non-singular theories and imposing that predictions are independent of where we place the cut-off is a consistency condition. In this sense, the complete theory would be defined when one takes the limit in which the cut-off is placed at infinity. In the more recent literature, the attitude is that even cut-offs (for renormalizable theories) can formally be sent to infinity, our theories are just effective theories valid up to certain scales\footnote{For a discussion of this perspective about the role of cut-offs in high-energy physics see \cite{Crowther2015} and references therein.}. 

%Let me also emphasize that while in the condensed matter case the renormalization group flow related different states that a system could have, in the high-energy case 

Closely related to the notion of renormalization group flow in high-energy physics there is the notion of running couplings. Other renormalization schemes depend not only on a cut-off scale $\Lambda$ or some similar regulator but on an additional renormalization scale $\mu$\footnote{See the discussion in \cite{Williams2018} and the references therein for other regularization methods.}. One can conceive of a renormalization group flow in which the scale $\mu$ changes and which implies a change in the parameters in the theory. This version of the flow is also a purely formal flow in the sense that it is a flow in a theory space defined by a series of couplings and the two scales which connects equivalent theories, i.e., theories that give the same predictions (e.g. same transition amplitudes) even if they are defined for different values of $\Lambda$ and $\mu$, which are considered just formal and unphysical parameters. However, if one sets the renormalization scale $\mu$ to be of the same order as some physical phenomenon (e.g. the center of mass energy of a collision), then the values of the couplings acquire a physical interpretation as they can be interpreted as the value of some physical properties like the mass or charge at a given scale. In this sense, the renormalization group flow can be given the interpretation of describing how the strength of certain interactions depends on the energy scale at which phenomena occur. In any case, the renormalization group flow in this context is still a (powerful) computational tool and hence it is just an abstract flow that doesn't represent a physical process. That is, it gives us information about a low-energy phenomenon given information about a high-energy one and this doesn't correspond to a physical process in the same sense that an ice cube melting is. The connection of this flow with phase transitions is not straightforward, although one could try to devise an argument connecting how the strength of the interactions change with energy scale with the connection between energy and temperature in condensed matter\footnote{For instance, in the case of matter as described by QCD one can build a phase diagram in which depending on the temperature and chemical potential there are different phases such as quark-gluon plasma or regular nuclear matter. In this context one could try to somehow heuristically connect this phase diagram with the flow of the renormalization group flow of the quantum field theory, as the temperature would be setting the relevant energy scales for renormalization.}.

Leaving aside these technicalities, let me insist that the renormalization group flow is straightforwardly connected with the phase structure of a system just in the condensed matter context, and that in any case it corresponds to some abstract flow and it is not describing some physical process\footnote{Of course, in the condensed matter case one could externally change the thermodynamic parameters of the system so that its physical evolution follows the renormalization group flow trajectory, but in this case the type of physical process we have is clearly the standard thermodynamical one and not something else related with renormalization.}. That is not to say that the renormalization group does not give us physical information, just that it is a computational tool and not representing a sequence of actual states that a given system has in different moments of its evolution. For the discussion of GFT, I will discuss in the next section that the particle physics and the condensed matter perspectives are sometimes mixed\footnote{To be fair, this issue can be argued to affect other presentations of the renormalization group in the physics literature.}, but in any case, the important message to take from this section is that physical phase transitions correspond to changes in the thermodynamic parameters of a system and these changes, as all changes between thermodynamic equilibrium states, are driven from outside the system\footnote{One could also consider a situation in which the thermodynamical parameters are parameters in the laws of physics and that they have a time-dependence, but one could still consider this time-dependence of the parameters as an external thing.}. Moreover, the way these parameters change depends on the environment of the system and it is completely independent of the renormalization group flow.

%Therefore we have to distinguish the renormalization group flow in the two contexts: it can either represent a coarse-graining process which gives us some information about the different phases of the system or it can represent an equivalence relation between high-energy theories from which .

% We will see later on that both types of flows seem to be mixed up in the discussion of GFT geometrogenesis. It is only the former flow which is straightforwardly relevant for the discussion of phase transitions, and it is important to keep in mind that this flow does not correspond to the evolution of a system undergoing a phase transition. 

%In this sense, let me insist that the renormalization group flow is just a formal flow which does not correspond to any physical process. Physical phase transitions correspond to changes of the thermodynamic parameters of a system and these changes, as all changes between thermodynamic equilibrium states, are driven from outside the system. Moreover, the way these parameters change depends on the environment of the system and it is completely independent of the renormalization group flow.

\section{Geometrogenesis in GFT} \label{Geometrogenesis_GFT}

Now we are in a position to analyze the idea of geometrogenesis as proposed in the GFT literature. This idea has been defended by Daniele Oriti in \cite{Oriti2014, Oriti2021}. As I have discussed in section \ref{GFT_condensates_cosmo}, condensate states are believed to represent cosmological spacetimes, once one identifies some variable like $\phi$ with a clock or time variable. \cite{Oriti2014} proposes two further hypotheses, that we conceive of a condensation process by which these condensate states are formed and that we identify this process with the big bang singularity. In this section I will analyze this proposal and I will find it problematic, as it seems to be assuming an additional and unwanted time dimension, even if Oriti tries to avoid obtaining it.

\cite{Oriti2014} introduces geometrogenesis in the following way:
\begin{quote}
That is, identify the process of quantum spacetime condensation with a known, even if not understood, physical process: the big bang singularity. Better, we identify the coming of the universe, that is of space and time, into being with the physical condensation of the “spacetime atoms”. There was no space and no time before this condensation happened. Therefore, we could call the spacetime condensation \textit{geometrogenesis}.
 \cite[p. 194, his emphasis]{Oriti2014}
\end{quote}
Notice that if we take seriously that there was no space and time `before' the condensation happened, we find some tension, as we would naively think of the condensation process as a process happening in time. Oriti adds a footnote to this passage precisely on this point:
\begin{quote}
The “tensed” wording is inevitable and can only refer to some internal time variable, which started running monotonically from the condensation onwards, just like the corresponding tensed statements about the evolution of the universe in (quantum) cosmology; for example, this variable could be a hydrodynamic variable corresponding to the volume of the universe.
\cite[Footnote 11, p. 194]{Oriti2014}
\end{quote} 
Here we see that Oriti is appealing to the notion of internal time (something like the relational time $\phi$ in my discussion above), which allows the definition of a relational evolution for a given state. However, as Oriti notes, this internal variable is only available from the condensation onwards, and hence it cannot work as a time variable for the transition itself. In this sense, this footnote does not clarify how we should take tensed language when referring to the transition and the way in which this transition is physical is so far unclear. 

Given these quotes and my discussion of the GFT formalism in this paper, it would seem reasonable to have a model of geometrogenesis like the following. Imagine that we had a solution of the GFT dynamical equation \ref{constraint_GFT} such that we could distinguish two `regions' or `parts' of the state. One of them would be the geometrical one which can be approximated by a spacetime and the other one would be the non-geometric one. If we took seriously the GFT formalism we could discuss this sort of geometrogenesis, which for sure would also be subject to deep conceptual troubles. However, Oriti does not seem to have this picture in mind, as the way he describes geometrogenesis is not as a consequence of the dynamics defined by equation \ref{constraint_GFT}, but as a phase transition analogous to the condensed matter phase transitions. In particular, Oriti explains the way the geometrogenesis phase transition is supposed to be implemented in the following way: 
\begin{quote}
We have seen that GFTs, just as the field theories describing the fundamental atoms in condensed matter systems, are defined usually in perturbative expansion around the Fock vacuum. In this approximation, they describe the interaction of quantized simplices and spin networks, in terms of spin foam models and simplicial gravity. The true ground state of the system, however, for non-zero couplings and for generic choices of the macroscopic parameters, will not be the Fock vacuum. The interacting system will organize itself around a new, non-trivial state, as we have seen in the case of standard Bose condensates. The relevant ground states (which, due to diffeomorphism invariance, cannot correspond to minima of an energy functional) for different values of the parameters (couplings, etc.) will correspond to the different macroscopic, continuum phases of the theory, with the dynamical transitions from one to the other being indeed phase transitions of the physical system we call spacetime. 
 \citep[p. 195]{Oriti2014} 
\end{quote}
In this passage Oriti makes the analogy with condensed matter systems which may have different phases for different values of the parameters which describe them. A key insight used in statistical mechanics to determine the macroscopic properties of a system is that, in equilibrium thermodynamic, states minimize or maximize certain thermodynamic potentials. For different values of the thermodynamic parameters, the thermodynamic potentials change, and this makes it the case that for different values of the parameters we may have different phases, which correspond to different extrema of the potentials. In the case of Bose condensates that Oriti mentions\footnote{See his presentation on section 5 of \cite{Oriti2014}.}, the thermodynamic potential that is minimized is directly the energy functional of the field theory associated with the condensate, and hence one is interested in the ground state of the system. A change in the parameters of the system may imply a change in the ground state of the system, leading to a phase transition. For a thermodynamic system, these parameters may be fixed, e.g., the masses of the molecules that form a gas, or may be externally changed, e.g., the temperature or pressure of the gas.

This analogy does not fit well with the case of GFT. First, in thermodynamics each state represents the way a system is at a given time, and we can represent a process in which this state changes as a process in time. However, in the case of GFT each condensate state is interpreted to represent a full spacetime (once we are interpreting $\phi$ as a time) and not a state of a system at a time\footnote{At the end of this section I will consider what would happen if we took a different `relational'  interpretation.}. If the system is spacetime as Oriti claims, a phase transition would require introducing an additional evolution parameter that would connect the different spacetimes or GFT states. This would amount to introducing a metatime or fifth dimension, which would be an undesired consequence. That is, evolution in this parameter would connect different spacetimes, and possibly also non-geometric GFT states which cannot be interpreted as spacetimes. But we lack any evidence of any such fifth dimension, and it is not clearly related to the big bang singularity. Hence we end up with some additional unwanted structure that does not do the job it was supposed to do. Not only do we lack information about this extra dimension, but we also lack information about the dynamics that would lead the parameters of the theory to change. Here we would find again duplicity: we would have the dynamics of the GFT theory as supposedly defined by \ref{constraint_GFT} and we would also have some additional dynamics, of which we don't know anything yet, that would drive the change of the parameters of the theory on this metatime. In the case of thermodynamics we can account for the change of the thermodynamic properties as changes external to the system, but in the case of GFT we seem to be lacking a story of how and why the parameters in the theory change and give rise to different phases. In this sense, the GFT phase transition, as a physical process, is unclear and somewhat mysterious as described in \cite{Oriti2014}.  

Let me also mention that from my discussion in the previous section it should be clear that we lack a clear understanding of the conditions that supposedly need to obtain in order to have condensate states being a solution to the GFT dynamics. In this sense, the claim that some values of the GFT parameters are associated with a condensate state and some of them are not is merely speculative, and more work in the foundations of GFT would be needed in order to build something like a phase diagram of a given GFT.

\cite{Oriti2021} addresses some of these issues. First, he rejects understanding the phase transition as a temporal process:
\begin{quote}
The main difficulty is the immediate temptation to interpret a cosmological phase transition not only as physical but also as a \textit{temporal} process. This is also a problem with the very language we use to characterize physical \textit{processes}. A phase transition is pictured as the outcome of ‘evolution’ in the phase diagram of the theory, or of a ‘flow’ of its coupling constants; we say we ‘move’ towards the cosmological, geometric phase from the non-geometric, non-spatiotemporal phase, or viceversa. However, we are dealing with a system which is already described at level 2: there is no continuum space, no continuum time, no geometry in the usual sense; and it is also not characterized by features which are just `one approximation away' from time and space.  \cite[p. 30, his emphasis]{Oriti2021}
\end{quote}
In this paper Oriti uses levels to speak about the way possible theories relate to continuum spacetime: the higher the level, the further away one is from the spacetime of general relativity. In this case, with level 2 he is referring to the phase which is described by a GFT state which cannot be given a geometric interpretation in terms of space and time. As at this level there is no notion of time, he rejects using temporal notions for describing the transition. However, if we are to take the analogy with the phase transitions of condensed matter systems seriously, we need to distinguish between the system, i.e., the GFT state/spacetime, and the transition itself. That the GFT state may not be spatiotemporal does not imply anything about the phase transition, of which we do not know much yet. In the 5-dimensional picture I was alluding to before, it seems plausible to ascribe a temporal meaning to the fifth dimension, even though there may already be some other notion of time in the model. 

In any case, Oriti continues:
\begin{quote}
So, first, we need to have a background-independent and non-spatiotemporal notion of ‘evolution’ in the space of quantum gravity coupling constants, i.e. in the ‘theory space’ characterizing the quantum gravity formalism at hand. Notice that such evolution will relate different continuum theories, in particular different macroscopic effective dynamics, for the same fundamental quantum entities. This notion of evolution in theory space is what specific renormalization group (RG) schemes in various quantum gravity formalism will provide.
\cite[pp. 30-31]{Oriti2021}
\end{quote} 
In this paragraph it is clear that for each set of parameters (with `coupling constants' it is referred to the parameters that may appear in the GFT action) one can associate a `theory', and hence a spacetime if we restrict ourselves to condensate states and interpret these states as spacetimes. Here Oriti introduces the flow of the renormalization group as providing a notion of evolution. As I have noted above, the renormalization group flow is generally not considered to be a physical flow associated with some physical process. In this passage it may seem that Oriti has the same picture in mind, given that he speaks about theory space and writes evolution in scare quotes. However, we know that the interpretation he wants to make of the flow and of transitions is to take them as physical processes. 

At this point it is important to make emphasis in that \cite{Oriti2021} isn't very explicit about which kind of renormalization group flow would be occurring in the case of GFT and about how to interpret it. Oriti cites some work in GFT \citep{Carrozza2014,Carrozza2016} in which the renormalizability of GFT is discussed. In these works, it seems that the high-energy physics perspective is adopted, that is, one introduces a regulator $\alpha$, similar to the cut-off $\lambda$ discussed in the previous section, to have mathematically well-defined quantities. The theories with different regulators $\alpha$ nevertheless agree on certain `infrared predictions', just as in QFT theories with different cut-offs make the same predictions in their shared ranges of validity. The complete theory would be obtained by completely eliminating the regulator $\alpha$, just as in the case of QFT a theory with a cut-off can be interpreted to be just an effective theory for certain energy regimes. In this sense, the perspective in \cite{Carrozza2016} about the renormalization group flow seems to be the high-energy physics perspective which isn't directly related to phase transitions. As discussed in the previous section, the change from a theory with a regulator $\alpha_1$ to another with regulator $\alpha_2$ just corresponds to a change in the mathematical structures we use for describing certain phenomena, but nothing about the phenomena. Therefore, this picture does not fit well with the condensed matter picture that Oriti is aiming for, as one can see in his use of the Bose-Einstein condensate example in both \cite{Oriti2014} and \cite{Oriti2021}. His discussion of Bose-Einstein condensation is the lines of the standard condensed matter picture of phase transitions: he mentions that under different conditions, the system will organize in different ways, corresponding to different phases. As I have argued above, this analogy does not work well for the GFT case, as it leads to a 5-dimensional picture. In any case, we find in Oriti and in the works in GFT intuitions from both applications of the renormalization group ideas, and it is not completely clear the way the renormalization group is to be implemented and understood. Furthermore, in neither case the flow of the renormalization group is generally understood to be a physical process in the sense of describing the changes that happen to a system, and taking this flow to be physical seems to imply an unwanted temporal dimension.

%In any case, even if one could define a renormalization group flow in GFT that would be closer to a zooming out process as in the condensed matter setting, the physicality of this process can be challenged on the same grounds as I exposed in the previous section and it wouldn't be able to overcome the challenge that it seems to be a process in an unwanted additional temporal dimension. 

\cite{Oriti2021} recognizes that there is a conceptual problem with the analogy with the condensed matter system:
\begin{quote}
The reason why we have no particular conceptual issue in understanding the flow in theory space and the approach to phase transitions in temporal terms, despite the fact that they refer to a change in the time-independent coupling constants of the system, is that we can easily imagine an external observer (the experimental physicist in the lab) tuning such coupling constants towards their critical values, and thus pushing the system towards the relevant phase transition. Needless to say, no such external observer is available in quantum gravity.
 \cite[p. 31]{Oriti2021}
\end{quote}
The phrasing here may suggest that the problem has to do with the notion of external observers, but the key point here, in my opinion, is that in the case of a condensed matter system we can think of the system as embedded in a spatiotemporal structure in which the thermodynamic parameters change, while in the case of the theory of quantum gravity the theory itself defines a spatiotemporal structure, and introducing a further process of change seems to introduce additional metatemporal dimension. Nevertheless, Oriti still claims that:
\begin{quote}
Any notion of time or, better, ‘proto-time’ that could be associated to such flow across the quantum gravity phase diagram would in any case deserve such name only in the sense that, once used to parametrize the flow across a non-geometric phase towards a geometrogenesis phase transition, it ends up matching some spatiotemporal observable that can be used as a time variable within the geometric phase. Viceversa, it would correspond to what is left of some geometric variable used to define a notion of time in such phase, and used as well as a notion of RG scale for the quantum gravity system, once the same system flows across a geometrogenesis phase transition into a non-spatiotemporal phase.
\cite[p. 31]{Oriti2021}
\end{quote}  
In this passage Oriti proposes that, in order to avoid having two times, one can identify time with proto-time, which is the way he calls the parameter which describes the renormalization group flow\footnote{This would presumably be something like the regulator $\alpha$ in models like \cite{Carrozza2016}.}. However, time and proto-time cannot be identified in any straightforward way: for any `instant' of proto-time we have a value for the coupling constants and, possibly, a full spacetime, with its whole range of possible times. In other words, the picture we have is clearly 5-dimensional, the two time parameters have nothing to do with one another, and there is no principled way to identify one with the other. Similarly, we have two dynamical processes: the one driving the change of the parameters in proto-time or metatime and the one defined by the GFT dynamical equation \ref{constraint_GFT}. Also note that the transition happens in proto-time but is not localized spatiotemporally, which means that, contrary to what we wanted, there is nothing associating the big bang singularity, which is localized in spacetime, with the phase transition.

In this sense, Oriti fails to provide a satisfactory answer to the challenge posed by the fact that adding an extra time parameter or proto-time makes our system become equivalent to a 5-dimensional one. Moreover, that the dynamics is given by the renormalization group flow seems ad hoc and it is unclear the way it is supposed to work. In this sense, it would be helpful for the proposal to have a concrete definition and justification of the mechanism by which the renormalization group flow is intended to act.

This discussion above was based on the assumption (shared in the earlier GFT papers\footnote{See the review \cite{Pithis2019} and the references therein.}) that if one relationally interprets GFT, states in the Hilbert space of the theory represent full spacetimes, and hence we found the extra time parameter problem. Alternative interpretations which also claim to be relational are also present in the literature\footnote{See for instance \cite{Oriti2014, Oriti2021}. I am thankful to an anonymous reviewer for letting me know about this interpretational shift.}, and even if they depart from the way relational dynamics is more broadly defined in the quantum mechanics and quantum cosmology literature, a potential benefit is that one could try to use this sort of interpretation to avoid the 5-dimensional problem, as one interprets states are interpreted as containing 3-dimensional information and not 4-dimensional one. However, even if we took this route we would find several serious issues that need at least clarification. For instance, one still seems to have two dynamics: the one defined by the equation \ref{constraint_GFT} or similar constructions\footnote{In works like \cite{Marchetti2021, Marchetti2021a} the way this dynamics is defined is slightly more complicated than what exposed above, as one defines dynamics as evolution with respect to a parameter $T$ relating a family of wavefunctions $\psi _T$ which satisfy equation \ref{constraint_GFT} or an equivalent version of it. This definition is not related in any evident way to any renormalization group flow.} and the one putatively associated with the renormalization group flow. As far as I can see, the claim that these two dynamics are the same seems to require at least further justification. Finally, we still have the problem that the renormalization group flow is generally not considered to be describing a physical process, and even in this changed picture one would need to account for this. In this sense, I conclude that changing to a different interpretation will not solve the serious issues found with the formulation of geometrodynamics. Finally, let me also mention that the original papers analyzed here by Oriti \cite{Oriti2014, Oriti2021} were previous to the different interpretation presented in \cite{Marchetti2021, Marchetti2021a}, so we can put into doubt that Oriti had that particular interpretation in mind. This further supports my claim that in the GFT context a clarification of how the theory is to be interpreted is necessary, as so far it leads to serious trouble.

Let me finish this section by comparing the idea of geometrogenesis as presented by Oriti with some other similar ideas present in the quantum gravity literature. I will note that when one discusses renormalization group flows in other approaches one risks falling into the same 5-dimensional problem if one interprets them as physical processes and that the idea of a phase transition can be implemented in a less troublesome way if one does not postulate an additional dynamics. 

Causal dynamical triangulations\footnote{See \cite{Loll2020} for a review of CDT and further references on the topic.} (CDT) is an approach that presents some connections with GFT, as it is based also on simplicial decompositions of spacetime. In the literature on this approach it is also spoken about the phase diagram of the theory and phase transitions, as depending on the values of the parameters of the theory different behaviors are to be expected. However, it is important to note that in this context the talk of phase transitions is, as far as I know, never explicitly taken as representing a physical process as Oriti takes phase transitions in GFT. If one took phase transitions as physical processes in the case of CDT one would run into the same problem as in the case of GFT, namely that it would require the addition of a fifth spacetime dimension connecting the discrete spacetimes that there are for each value of the parameters. The other sense in which it is interesting to study the phase diagram of CDT and its behavior under the renormalization group, apart from classifying the different ways the system, i.e., spacetime can behave, is that this study can give us information about the continuum limit of the theory. As I explained in the previous section, the action of the renormalization group can be roughly understood as a process of zooming out, and hence it is useful for obtaining knowledge of a theory from the point of view of large scales. In the context of CDT it is hoped that the renormalization group gives some information about how to obtain a theory in which the discretization is eliminated. In conclusion, even if in CDT phase transitions are mentioned, it is just generally in a formal way and not representing a physical process like the GFT geometrogenesis that Oriti has in mind.

The original idea of geometrogenesis was developed in the context of quantum graphity\footnote{See \cite{Konopka2006} for a technical overview of the approach and \cite{Markopoulou2009} for a more conceptual discussion of the approach and of the idea of geometrogenesis.}, a different approach to quantum gravity. In this approach one has a 3+1 split, which makes the idea of geometrogenesis more intelligible, even if perhaps less radical. Quantum graphity describes the quantum evolution of a graph formed by $N$ nodes that can be connected or not. When the nodes are connected in a regular and appropriate way they approximate a space, just as I have argued that some states in LQG and GFT approximate a space. Some other states are certainly not approximations to spaces, such as the state in which every node is connected to every other node. In quantum graphity one describes the quantum evolution of this system given a Hamiltonian. In this context, if by starting with a non-spatial state the evolution of the system leads to a state which approximates a space, then one can speak about geometrogenesis. In the context of GFT one could try to mimic this simpler model of geometrogenesis if one were able to find a 3+1 split or a definition of the dynamics with respect to an external time parameter.

Finally, there are some models with transitions that seem more similar to what Oriti was aiming for. For instance, consider the class of loop quantum cosmology models studied in \cite{Huggett2018}. In these models there is a sense in which one can argue that a non-spatiotemporal region happened `before' our spatiotemporal region. In particular, there is a signature change, that is, the model is based on a manifold that has some regions with Lorentzian signature and some regions with Euclidean signature. Arguably, one can claim that there is time only in the regions with Lorentzian signature, and that in `going' from some region to another time appears or disappears. This would have happened at the big bang: there was a non-spatiotemporal region of the universe `contiguous' to our spatiotemporal region\footnote{To clarify: Huggett and W\"uthrich's discussion is clearest using the signature change example in which one has an underlying manifold, but they allow for other possibilities. That is, there can be something like a geometrogenesis as long as there are some parts or regions of the fundamental quantum gravity stuff, whatever it is, which approximate spacetimes and some that do not.}. This kind of picture seems to be what Oriti aimed for and what I have commented above that could be argued for in the context of GFT if we had states which could be decomposed into geometric and non-geometric parts. This is in contrast with the 5-dimensional picture that Oriti reaches, although it is also subject to important conceptual worries, given that the way GFT states are to be interpreted can be challenged, as I have pointed out in this paper.

In this sense, the analysis of similar ideas in other approaches illustrates the main conceptual difficulty of the proposal of geometrogenesis as proposed by Oriti. In Oriti's idea of geometrogenesis, which could be applied to models like CDT, one ends up postulating additional dynamics in an extra temporal dimension, which is unwanted as GFT and CDT models already had a dynamics and a temporal dimension (even if perhaps only for certain states). The ways geometrogenesis or phase transitions have been formulated in quantum graphity or LQC are conceptually less problematic, as they are based on the dynamics of these approaches and do not need to add any extra dimension or dynamics. In this sense, these ideas seem more promising for implementing the geometrogenesis picture that Oriti is aiming for.

\section{Conclusions}\label{conclusions}

In this paper I have analyzed the idea of geometrogenesis as it has been introduced in the GFT literature. I have found that the way it is introduced is problematic, as it requires the incorporation of some additional and unwanted temporal dimension and also the postulation of an extra layer of dynamics. This would get in conflict with the way time and dynamics are defined in GFT in general and in its cosmological models in particular. Even if we move to an alternative, more exotic interpretation of the GFT formalism, I have argued that the deep issues highlighted in this paper need to be addressed. In particular, the conclusion one reaches by analyzing the proposal of geometrogenesis is that the foundations of GFT need to face several issues such as how to provide a satisfactory interpretation of the formalism, and that the introduction of the renormalization group flow into the discussion is clearly problematic and not properly justified.

As an alternative, it seems more promising to base the idea of geometrogenesis on the states and dynamics of the theory. That is, to look for solutions to the dynamical equations of the theory that would allow to identify spatiotemporal and non-spatiotemporal regions. This would put geometrogenesis in the GFT context in a similar position to other similar proposals I have mentioned in this paper, although there would remain certain concerns, both about the idea of geometrogenesis in general and about the intelligibility and feasibility of GFT as a physical theory.

\section*{Acknowledgments}
I want to thank the Proteus group, Carl Hoefer, and, especially, Daniele Oriti for their comments and discussions. This research is part of the Proteus project that has received funding from the European Research Council (ERC) under the  Horizon 2020 research and innovation programme (Grant agreement No. 758145) and of the project CHRONOS (PID2019-108762GB-I00) of the Spanish Ministry of Science and Innovation.

\bibliographystyle{chicago}% bibliography style

\bibliography{bibliography.bib}

\begin{thebibliography}{}

\bibitem[\protect\citeauthoryear{Anderson}{Anderson}{2017}]{Anderson2017}
Anderson, E. (2017).
\newblock {\em {The Problem of Time}}, Volume 190 of {\em Fundamental Theories
  of Physics}.
\newblock Cham: Springer International Publishing.

\bibitem[\protect\citeauthoryear{Butterfield and Bouatta}{Butterfield and
  Bouatta}{2016}]{Butterfield2014a}
Butterfield, J. and N.~Bouatta (2016, jan).
\newblock {Renormalization for Philosophers}.
\newblock In {\em Metaphysics in Contemporary Physics}, pp.\  437--485. BRILL.

\bibitem[\protect\citeauthoryear{Cardy}{Cardy}{1996}]{Cardy2015}
Cardy, J. (1996, apr).
\newblock {\em {Scaling and Renormalization in Statistical Physics}}.
\newblock Cambridge University Press.

\bibitem[\protect\citeauthoryear{Carrozza}{Carrozza}{2016}]{Carrozza2016}
Carrozza, S. (2016).
\newblock {Flowing in group field theory space: A review}.

\bibitem[\protect\citeauthoryear{Carrozza, Oriti, and Rivasseau}{Carrozza
  et~al.}{2014}]{Carrozza2014}
Carrozza, S., D.~Oriti, and V.~Rivasseau (2014).
\newblock {Renormalization of Tensorial Group Field Theories: Abelian U(1)
  Models in Four Dimensions}.
\newblock {\em Communications in Mathematical Physics\/}~{\em 327\/}(2).

\bibitem[\protect\citeauthoryear{Chirco, Kotecha, and Oriti}{Chirco
  et~al.}{2019}]{Chirco2019}
Chirco, G., I.~Kotecha, and D.~Oriti (2019).
\newblock {Statistical equilibrium of tetrahedra from maximum entropy
  principle}.
\newblock {\em Physical Review D\/}~{\em 99\/}(8), 1--18.

\bibitem[\protect\citeauthoryear{Chua and Callender}{Chua and
  Callender}{2021}]{Chua2021}
Chua, E.~Y. and C.~Callender (2021, may).
\newblock {No time for time from no-time}.
\newblock {\em Philosophy of Science\/}~{\em 88\/}(5), 1172----1184.

\bibitem[\protect\citeauthoryear{Crowther}{Crowther}{2015}]{Crowther2015}
Crowther, K. (2015).
\newblock {Decoupling emergence and reduction in physics}.
\newblock {\em European Journal for Philosophy of Science\/}~{\em 5\/}(3),
  419--445.

\bibitem[\protect\citeauthoryear{Freidel}{Freidel}{2005}]{Freidel2005}
Freidel, L. (2005, oct).
\newblock {Group field theory: An overview}.
\newblock {\em International Journal of Theoretical Physics\/}~{\em 44\/}(10),
  1769--1783.

\bibitem[\protect\citeauthoryear{Gabbanelli and {De Bianchi}}{Gabbanelli and
  {De Bianchi}}{2020}]{Gabbanelli2020}
Gabbanelli, L. and S.~{De Bianchi} (2020, aug).
\newblock {Cosmological implications of the hydrodynamical phase of group field
  theory}.
\newblock {\em General Relativity and Gravitation\/}~{\em 53\/}(7).

\bibitem[\protect\citeauthoryear{Gielen}{Gielen}{2018}]{Gielen2018}
Gielen, S. (2018).
\newblock {Group field theory and its cosmology in a matter reference frame}.
\newblock {\em Universe\/}~{\em 4\/}(10), 1--18.

\bibitem[\protect\citeauthoryear{Gielen}{Gielen}{2019}]{Gielen2019}
Gielen, S. (2019).
\newblock {Inhomogeneous universe from group field theory condensate}.
\newblock {\em Journal of Cosmology and Astroparticle Physics\/}~{\em
  2019\/}(2), 1--25.

\bibitem[\protect\citeauthoryear{Gielen, Oriti, and Sindoni}{Gielen
  et~al.}{2013}]{Gielen2013a}
Gielen, S., D.~Oriti, and L.~Sindoni (2013, nov).
\newblock {Homogeneous cosmologies as group field theory condensates}.
\newblock {\em Journal of High Energy Physics\/}~{\em 2014\/}(6).

\bibitem[\protect\citeauthoryear{Gielen and Sindoni}{Gielen and
  Sindoni}{2016}]{Gielen2016}
Gielen, S. and L.~Sindoni (2016).
\newblock {Quantum cosmology from group field theory condensates: A review}.
\newblock {\em Symmetry, Integrability and Geometry: Methods and Applications
  (SIGMA)\/}~{\em 12}.

\bibitem[\protect\citeauthoryear{Goldenfeld}{Goldenfeld}{1992}]{Goldenfeld1992}
Goldenfeld, N. (1992, mar).
\newblock {\em {Lectures on Phase Transitions and the Renormalization Group}}.
\newblock CRC Press.

\bibitem[\protect\citeauthoryear{Gryb and Th{\'{e}}bault}{Gryb and
  Th{\'{e}}bault}{2016}]{Gryb2016}
Gryb, S. and K.~P. Th{\'{e}}bault (2016).
\newblock {Time remains}.
\newblock {\em British Journal for the Philosophy of Science\/}~{\em 67\/}(3),
  663--705.

\bibitem[\protect\citeauthoryear{H{\"{o}}hn, Smith, and Lock}{H{\"{o}}hn
  et~al.}{2021}]{Hohn2021}
H{\"{o}}hn, P.~A., A.~R. Smith, and M.~P. Lock (2021, sep).
\newblock {Trinity of relational quantum dynamics}.
\newblock {\em Physical Review D\/}~{\em 104\/}(6), 066001.

\bibitem[\protect\citeauthoryear{Huggett and W{\"{u}}thrich}{Huggett and
  W{\"{u}}thrich}{2018}]{Huggett2018}
Huggett, N. and C.~W{\"{u}}thrich (2018, dec).
\newblock {The (A)temporal Emergence of Spacetime}.
\newblock {\em Philosophy of Science\/}~{\em 85\/}(5), 1190--1203.

\bibitem[\protect\citeauthoryear{Isham}{Isham}{1993}]{Isham1993}
Isham, C.~J. (1993).
\newblock {Canonical Quantum Gravity and the Problem of Time}.
\newblock {\em Integrable Systems, Quantum Groups, and Quantum Field
  Theories\/}, 157--287.

\bibitem[\protect\citeauthoryear{Kiefer}{Kiefer}{2012}]{Kiefer2012}
Kiefer, C. (2012, apr).
\newblock {\em {Quantum Gravity}\/} (Third ed.).
\newblock New York, NY: Oxford University Press.

\bibitem[\protect\citeauthoryear{Konopka, Markopoulou, and Smolin}{Konopka
  et~al.}{2006}]{Konopka2006}
Konopka, T., F.~Markopoulou, and L.~Smolin (2006, nov).
\newblock {Quantum Graphity}.

\bibitem[\protect\citeauthoryear{Kuchař}{Kuchař}{1992}]{Kuchar1992}
Kuchař, K.~V. (1992, jul).
\newblock {Time and interpretations of quantum gravity}.
\newblock In G.~Kunstatter, D.~Vincent, and J.~Williams (Eds.), {\em
  Proceedings of the 4th Canadian Conference on General Relativity and
  Relativistic Astrophysics}, Singapore. World Scientific Publishing Company.

\bibitem[\protect\citeauthoryear{Kumar}{Kumar}{1993}]{Binney1992}
Kumar, P. (1993).
\newblock {The theory of critical phenomena: An introduction to the
  renormalization group. By J. J. Binney, N. J. Dowrick, A. J. Fisher, and M.
  E. J. Newman, Clarendon Press, Oxford, 1992. 464 pp.}
\newblock {\em International Journal of Quantum Chemistry\/}~{\em 46\/}(5),
  671--671.

\bibitem[\protect\citeauthoryear{Loll}{Loll}{2020}]{Loll2020}
Loll, R. (2020, may).
\newblock {Quantum gravity from causal dynamical triangulations: A review}.
\newblock {\em Classical and Quantum Gravity\/}~{\em 37\/}(1).

\bibitem[\protect\citeauthoryear{Marchetti and Oriti}{Marchetti and
  Oriti}{2021a}]{Marchetti2021}
Marchetti, L. and D.~Oriti (2021a, may).
\newblock {Effective relational cosmological dynamics from quantum gravity}.
\newblock {\em Journal of High Energy Physics\/}~{\em 2021\/}(5), 25.

\bibitem[\protect\citeauthoryear{Marchetti and Oriti}{Marchetti and
  Oriti}{2021b}]{Marchetti2021a}
Marchetti, L. and D.~Oriti (2021b, jul).
\newblock {Quantum Fluctuations in the Effective Relational GFT Cosmology}.
\newblock {\em Frontiers in Astronomy and Space Sciences\/}~{\em 8}, 110.

\bibitem[\protect\citeauthoryear{Markopoulou}{Markopoulou}{2009}]{Markopoulou2009}
Markopoulou, F. (2009, sep).
\newblock {Space does not exist, so time can}.

\bibitem[\protect\citeauthoryear{Menon and Callender}{Menon and
  Callender}{2013}]{Menon2013}
Menon, T. and C.~Callender (2013, feb).
\newblock {Turn and Face the Strange... Ch-Ch-Changes: Philosophical Questions
  Raised by Phase Transitions}.
\newblock In {\em The Oxford Handbook of Philosophy of Physics}, pp.\
  189--223.

\bibitem[\protect\citeauthoryear{Mielczarek}{Mielczarek}{2017}]{Mielczarek2017}
Mielczarek, J. (2017).
\newblock {Big Bang as a Critical Point}.
\newblock {\em Advances in High Energy Physics\/}~{\em 2017}, 4015145.

\bibitem[\protect\citeauthoryear{{Mozota Frauca}}{{Mozota
  Frauca}}{2023}]{MozotaFrauca2023}
{Mozota Frauca}, {\'{A}}. (2023, jan).
\newblock {Reassessing the problem of time of quantum gravity}.
\newblock {\em General Relativity and Gravitation\/}~{\em 55\/}(1), 21.

\bibitem[\protect\citeauthoryear{Oriti}{Oriti}{2007}]{Oriti2007}
Oriti, D. (2007, oct).
\newblock {Group field theory as the microscopic description of the quantum
  spacetime fluid: A new perspective on the continuum in quantum gravity}.
\newblock In {\em Proceedings of Science}.

\bibitem[\protect\citeauthoryear{Oriti}{Oriti}{2012}]{Oriti2012}
Oriti, D. (2012).
\newblock {The microscopic dynamics of quantum space as a group field theory}.
\newblock {\em Foundations of Space and Time: Reflections on Quantum
  Gravity\/}~{\em 9780521114}, 257--320.

\bibitem[\protect\citeauthoryear{Oriti}{Oriti}{2014}]{Oriti2014}
Oriti, D. (2014).
\newblock {Disappearance and emergence of space and time in quantum gravity}.
\newblock {\em Studies in History and Philosophy of Science Part B - Studies in
  History and Philosophy of Modern Physics\/}~{\em 46\/}(1), 186--199.

\bibitem[\protect\citeauthoryear{Oriti}{Oriti}{2016}]{Oriti2016b}
Oriti, D. (2016).
\newblock {Group field theory as the second quantization of loop quantum
  gravity}.
\newblock {\em Classical and Quantum Gravity\/}~{\em 33\/}(8), 1--24.

\bibitem[\protect\citeauthoryear{Oriti}{Oriti}{2017}]{Oriti2017}
Oriti, D. (2017).
\newblock {The universe as a quantum gravity condensate}.
\newblock {\em Comptes Rendus Physique\/}~{\em 18\/}(3-4), 235--245.

\bibitem[\protect\citeauthoryear{Oriti}{Oriti}{2021}]{Oriti2021}
Oriti, D. (2021, aug).
\newblock {Levels of Spacetime Emergence in Quantum Gravity}.
\newblock In C.~Wuthrich, B.~{Le Bihan}, and N.~Hugget (Eds.), {\em Philosophy
  Beyond Spacetime}, pp.\  16--40. Oxford University Press.

\bibitem[\protect\citeauthoryear{Oriti, Sindoni, and Wilson-Ewing}{Oriti
  et~al.}{2016}]{Oriti2016a}
Oriti, D., L.~Sindoni, and E.~Wilson-Ewing (2016).
\newblock {Emergent Friedmann dynamics with a quantum bounce from quantum
  gravity condensates}.
\newblock {\em Classical and Quantum Gravity\/}~{\em 33\/}(22).

\bibitem[\protect\citeauthoryear{Pithis and Sakellariadou}{Pithis and
  Sakellariadou}{2019}]{Pithis2019}
Pithis, A.~G. and M.~Sakellariadou (2019).
\newblock {Group field theory condensate cosmology: An appetizer}.
\newblock {\em Universe\/}~{\em 5\/}(6).

\bibitem[\protect\citeauthoryear{Reisenberger and Rovelli}{Reisenberger and
  Rovelli}{2001}]{Reisenberger2001}
Reisenberger, M.~P. and C.~Rovelli (2001, jan).
\newblock {Spacetime as a Feynman diagram: The connection formulation}.
\newblock {\em Classical and Quantum Gravity\/}~{\em 18\/}(1), 121--140.

\bibitem[\protect\citeauthoryear{Rogel-Salazar}{Rogel-Salazar}{2013}]{Rogel-Salazar2013}
Rogel-Salazar, J. (2013, jan).
\newblock {The Gross-Pitaevskii equation and Bose-Einstein condensates}.
\newblock {\em European Journal of Physics\/}~{\em 34\/}(2), 247--257.

\bibitem[\protect\citeauthoryear{Rovelli}{Rovelli}{1991}]{Rovelli1991a}
Rovelli, C. (1991, jan).
\newblock {Time in quantum gravity: An hypothesis}.
\newblock {\em Physical Review D\/}~{\em 43\/}(2), 442.

\bibitem[\protect\citeauthoryear{Rovelli}{Rovelli}{2004}]{Rovelli2004}
Rovelli, C. (2004, nov).
\newblock {\em {Quantum Gravity}}.
\newblock Cambridge: Cambridge University Press.

\bibitem[\protect\citeauthoryear{Rovelli and Vidotto}{Rovelli and
  Vidotto}{2015}]{Rovelli2015}
Rovelli, C. and F.~Vidotto (2015, jan).
\newblock {\em {Covariant loop quantum gravity: An elementary introduction to
  quantum gravity and spinfoam theory}}.
\newblock Cambridge: Cambridge University Press.

\bibitem[\protect\citeauthoryear{Williams}{Williams}{2018}]{Williams2018}
Williams, P. (2018).
\newblock {Renormalization Group Methods}.
\newblock In E.~Knox and A.~Wilson (Eds.), {\em The Routledge Companion to the
  Philosophy of Physics}. Routledge.

\bibitem[\protect\citeauthoryear{Zee}{Zee}{2010}]{Zee2010}
Zee, A. (2010).
\newblock {\em {Quantum field theory in a nutshell}}.
\newblock Princeton University Press.

\end{thebibliography}

\end{document}